\documentclass[
 twocolumn,
 amsmath,amssymb,
 aps, pre
]{revtex4-2}
\usepackage[utf8]{inputenc}
\DeclareUnicodeCharacter{2009}{\,}
\usepackage{graphicx}
\usepackage{dcolumn}
\usepackage{bm}
\usepackage{hyperref}
\usepackage{makecell}
\usepackage{caption}
\hypersetup{
    colorlinks=true,
    linkcolor=blue,     
    citecolor=blue,     
    urlcolor=blue       
}
\usepackage{physics}
\usepackage[
margin=1in,
]{geometry}
\usepackage{float}
\usepackage{xcolor}
\begin{document}
\preprint{APS/123-QED}
\title{\textbf{Dynamics of Marangoni-Driven Elliptical Janus Particles} 
}
\author{Pabitra Masanta}

\thanks{These authors contributed equally}
\author{Ratan Sarkar}
\thanks{These authors contributed equally}
\author{P. Parmananda}
\author{Raghunath Chelakkot}
\email{raghu@phy.iitb.ac.in}
\affiliation{%
 Department of Physics, Indian Institute of Technology Bombay, Mumbai, India\\
}%



\begin{abstract}
We investigate the spontaneous motion of an elliptical Janus particle, driven by Marangoni forces,
on a water surface to understand how particle shape and size influence its dynamics. The Janus
particle is one-half infused with a substance such as camphor, which lowers the surface tension
upon release onto the water surface. The resulting surface tension gradient generates Marangoni
forces that propel the particle. For fully camphor-infused (non-Janus) particles, previous studies
have shown that motion occurs along the short axis of the ellipse. However, for Janus particles, our
experiments reveal a much richer steady-state dynamics, depending on both the particle’s
eccentricity and size.
To understand these dynamics, we develop a numerical model that captures the connection between
the spatio-temporal evolution of the camphor concentration field and the Marangoni force driving
the particle. Using this model, we simulate the motion of particles with varying eccentricities—
from nearly circular to highly elongated shapes. The simulations qualitatively reproduce all the
trajectories observed in experiments and provide insights into how particle geometry influences the
dynamics of chemically driven anisotropic particles. With the help of the numerical model, we compute a full phase diagram characterising the dynamical states as a function of surfactant properties.
\end{abstract}

\maketitle

\newpage

\section{\label{sec:Intro}Introduction\protect\\}
The generation of spontaneous motion is a key feature in active matter systems. In nature, microbial swimmers such as E. coli propel themselves by rotating or beating their flagella~\cite{elgeti2015physics}. Other species, such as M. Xanthus, glide on soft hydrogel substrates using a combination of mechanisms other than swimming~\cite{gloag2013self}. In addition, a thin layer at the air-water interface (aquatic surface monolayers), which has distinct physicochemical properties, accommodates a rich ecosystem for microbes and insects that move on the interface~\cite{cunliffe2011microbiology,bush2006walking}. 
Inspired by these naturally occurring self-motile systems, a large class of artificially motile colloidal systems has been synthesised by various means~\cite{Therkauff2012, Palacci2013, Volpe2011, Buttinoni2013, Bricard2013, Thutupalli2011, Bechinger2016, vanderLinden2019, Nishiguchi2015, Yan2016}. One of the common ways to realise chemically active colloidal particles is by using Janus colloidal particles, with unequal chemical properties that provide a head-tail asymmetry~\cite{bechinger2016active, Therkauff2012, Volpe2011}. These unequal chemical properties create local variations in temperature or chemical potential, which induce local flows and result in a directed motion of the particle. 

In the recent past, the surfactant-water system has served as a prototype of such chemo-mechanical transducers, capable of providing directed motion of suspended particles. In these systems, spontaneous motion is generated due to Marangoni flow due to surface tension gradients, generated by adding chemicals that temporarily reduce surface tension~\cite {pimienta2014self} and such Marangoni-driven self-propulsion is achieved by a variety of mechanisms~\cite{kwak2021marangoni, hayashima2001camphor, nagai2005mode, Nakata01, lagzi2010maze, toyota2009self, nakata_collective_inanimate_boats, okamoto}. Among the wide variety of Marangoni propellers, one of the most common and extensively studied systems exhibiting spontaneous motion is the camphor–water system. The self-propulsion of camphor on water has fascinated researchers for decades, with the earliest studies dating back more than two centuries~\cite{tomlinson1862ii}. More recently, a series of studies on camphor systems have displayed a variety of collective phenomena, including rotational, translational, and oscillatory movements~\cite{hayashima2001camphor, hayashima, nakata, Nakata01, kohira,nakata1998,nakata2000,suematsu2010,nakata2015,suematsu2014quantitative, Nakata01, path_selection,AKELLA20181176}. To complement these experiments, theoretical analyses have also been conducted on such systems~\cite{Kitahata2}.

In these experiments, typically, a small paper disk infused with camphor is placed on the surface of water. As camphor molecules diffuse over the interface, they gradually reduce the local surface tension. A slight asymmetry in the camphor concentration, usually caused by a random perturbation, breaks the symmetry of the surface tension gradient and sets the disk in motion along a particular direction. The movement of the disk acts as a positive feedback on the asymmetric distribution of camphor, leading to a comet-like camphor profile around the moving disk. Single-particle\cite{Nakata_dancing, RandT_camphor} and collective dynamics of circular disks have exhibited oscillations\cite{ishant_filament}, synchronisations\cite{ribbon_synchronization, sharma2020rotational, Chimeralike, Nakata_synchro, liberation_motion, jain2023phase}, aperiodic bursts~\cite{sharma2022aperiodic}, kinetic arrest, and cluster flow of the particles. On the other hand, the polar symmetry of camphor concentration can be broken by the geometry of the paper disk. For instance, experiments on elliptical particles, completely dipped in camphor, have shown directed motion always in the direction of their short axis~\cite{kitahata}. Another effective way to break the polar symmetry of the camphor disk is to infuse camphor only on a portion of the disk,  rather than coating it completely. This can be done systematically by making Janus camphor disks of circular and elliptical shapes, by coating only one side of their symmetric axis.

Here, we experimentally study the dynamics of such an elliptical and circular Janus particle on the air-water interface, created by an asymmetric coating of camphor on the disk. The experiments show that the qualitative features of the particle are completely different compared to the non-Janus system, and their dynamics depend on the shape (measured by the eccentricity) of the Janus particle. Furthermore, we built a minimal theoretical model that captures the essential features of particle dynamics. This model takes into account the dynamics of the camphor molecule, as well as the particle. We numerically solve this model and systematically study the dynamics of the Janus particle by varying the particle shape, size, and the specific properties of the surfactants.

\section{\label{sec:level1}Experiments:\protect\\}
\subsection{ Method and preparation}
 \begin{figure}[H]
    \centering
     \includegraphics[width=.8\linewidth]{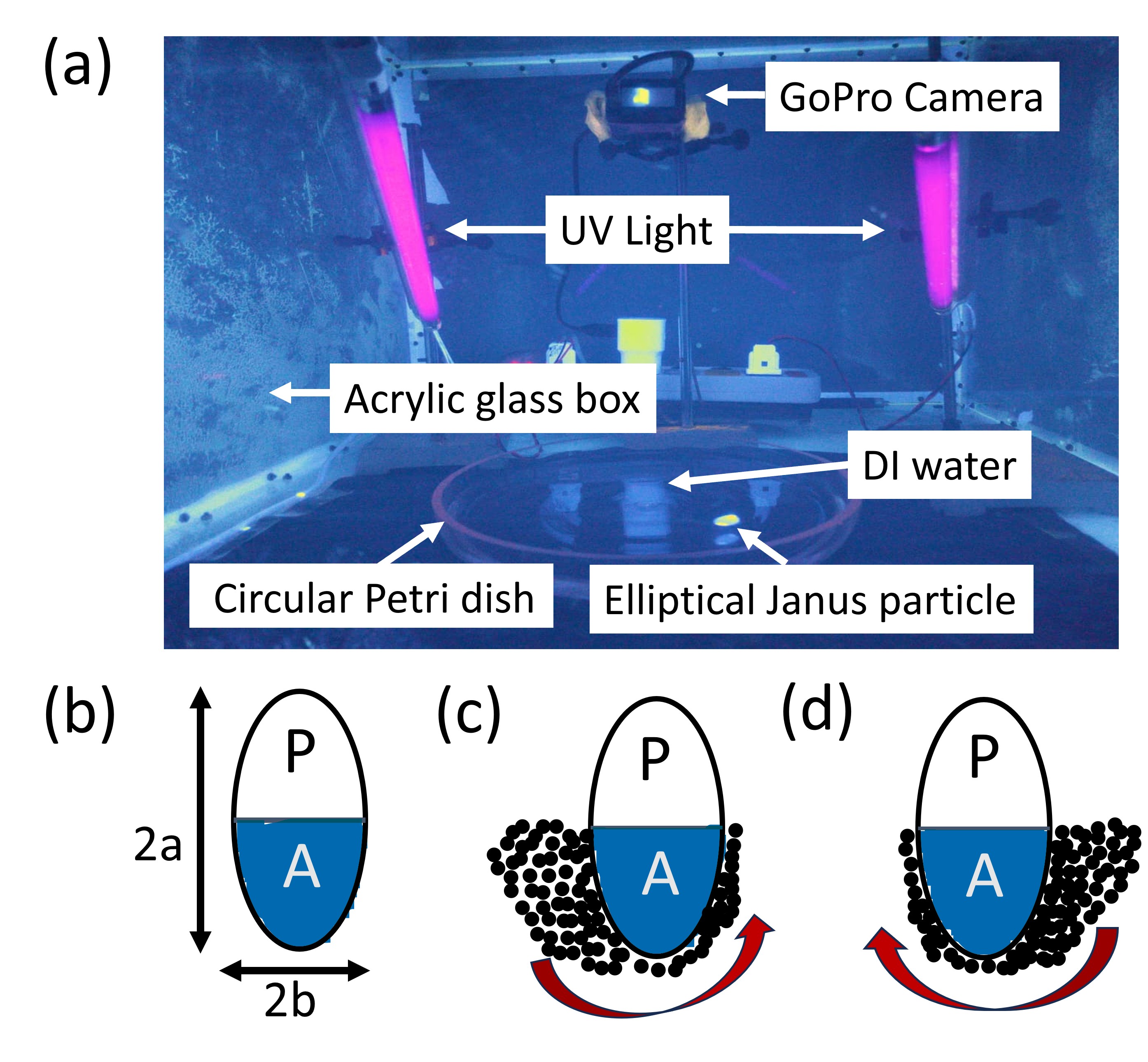}
     \caption{(a) Experimental setup. (b) Schematic of an elliptical Janus particle with major axis 2a and minor axis 2b. (c) Anti-clockwise rotation of the Janus particle. (d) Clockwise rotation of the Janus particle. (In Figs. (b), (c) and (d), A denotes the camphor-infused active part and P the passive part. In Figs. (c) and (d), the black dots represent the released camphor molecules.)}
     \label{fig:1}
 \end{figure}

For the experiments, Janus-like particles were prepared from A4 sheet paper by drawing ellipses of varying eccentricities. The length of the major axis 2a was fixed at 1.9 cm, while the length of the minor axis 2b was systematically decreased, starting from 2b= 1.9 cm, 1.6 cm, 1.4 cm, 0.8 cm and 0.6 cm, as illustrated in FIG.\ref{fig:2} (a) - (e) To create Janus-like particles, one half of each ellipse (the half separated by the minor axis, as illustrated in FIG.\ref{fig:1}.(b)) was painted with a 0.5 M camphor-ethanol solution using an acrylic marker pen\cite{song2023pen}. After painting, the particles were allowed to dry for about 2 minutes to enable evaporation of ethanol, leaving camphor deposited only on the painted half of the particle surface. Each particle was then gently placed approximately at the center on the surface of deionized (DI) water contained in a circular Petri dish with diameter, D= 26.5 cm. The Petri dish was filled with DI water of volume, V= 600 ml, creating a water height of about 2 cm. Separate experiments were conducted for each particle size mentioned above.

The entire experimental set-up was operated under UV light illumination to enhance visualization, since the two ends of the elliptical particles contain white circular spots that reflected UV light effectively. To minimize air interference and ensure accurate dynamics, the system was enclosed in a large acrylic glass box (see FIG.\ref{fig:1}). Room temperature was maintained at $(23 \pm 1)^0 $ C throughout all experiments. Particle dynamics on the water surface were recorded using a GoPro camera set to 60 frames per second and 1080p resolution. The captured videos were analysed using IMAGEJ software, while further analysis, such as estimation of trajectories and angular velocities, was performed with Python programming.

\subsection{Experimental results}
\begin{figure}
    \includegraphics[width=1.0
    \linewidth]{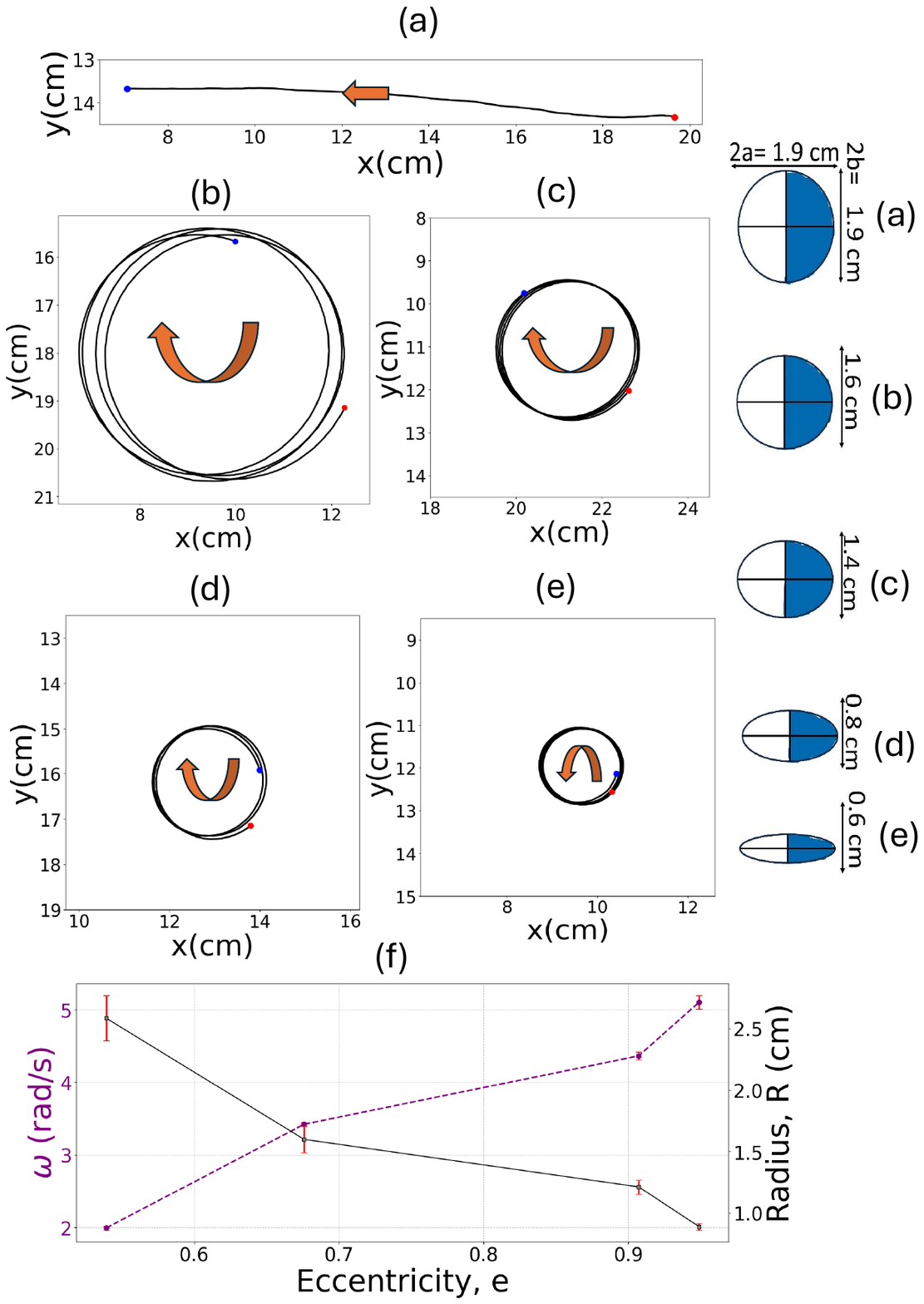}
     \caption{(a) - (e) Trajectory of centre of mass of  Janus particle for different shapes. 
(a) shows a straight path, while (b)--(e) show almost circular trajectories with radius $R$.  (b)~$R = (2.585 \pm 0.184)~\mathrm{cm}$; 
(c) $R = (1.599 \pm 0.105)~\mathrm{cm}$; 
(d) $R = (1.213 \pm 0.056)~\mathrm{cm}$; 
(e) $R = (0.888 \pm 0.028)~\mathrm{cm}$; 
(See Movie1) (f) shows the variation of angular speed $\omega$ and radius $R$ with eccentricity $e$. The average $\omega$ and $R$ are computed over ten cycles.}
\label{fig:2}
 \end{figure}

When the Janus particle is placed on the surface of the water, it starts to move at the air-water interface. This motion arises due to the fluid flow induced by the solutal Marangoni effect\cite{Nakata01}, which results from a surface tension gradient created by the camphor coating applied to one half of the particle. Depending on the particle size and shape, the Janus particles exhibit two distinct modes of motion. The circular Janus particles ($a \simeq b$) move in a straight path on the water surface, whereas the elliptical Janus particles display circular trajectories which are also periodic. In our experiments, the major axis was kept fixed at $2a = 1.9~\mathrm{cm}$, while the minor axis $2b$ was varied as $1.6$, $1.4$, $0.8$, and $0.6~\mathrm{cm}$. { For perfect circular particles, ( 2a = 2b= 1.9 cm), the particle moves in a straight line (Fig~\ref{fig:2}(a)), until the motion gets interrupted by the edges of the petri dish. However, when $2b<2a$, the particle follows a circular path as shown in Fig~\ref{fig:2}(b)-(e)}. We also observe that as $b$ decreased, the particle's radius of the circular paths decreased monotonously, as shown in Fig.\ref{fig:2}(b)-(e). In Fig.~\ref{fig:2}(f) the mean angular velocity $\omega$ and radius of the particle trajectories are calculated. 
It is evident that as the radius of the circle decreases, the angular speed of elliptical Janus particle increases (Fig.\ref{fig:2}(f)). At the beginning of each experiment, the particle exhibits high activity and speed, often causing it to collide with the container boundary. Therefore, we neglected this transient phase (about 60 seconds) of motion and analyzed only the portion of the experiments in which the particle's motion became stable and far from the boundary. To achieve such stable behavior, we used a camphor solution of 0.5 M instead of a higher molarity camphor solution, as a lower molarity allows the particle to reach a steady motion relatively quickly. Also, being a finite system, the particle dynamics gradually slows down. Therefore, the averages are calculated over  a time-window where such variations are negligible. 
In the case of a circular Janus particle, its symmetry ensures that the net force generated by the active region acts through the center-of-mass of the particle. As a result, the particle moves along an almost straight trajectory (Fig.\ref{fig:2}.a), although the release of camphor molecules is stochastic in nature. However, for an elliptical Janus particle, the stochastic release of camphor molecules causes the net force to deviate from the center-of-mass, producing a torque that induces rotation.
As the minor axis 2b decreases, the variation from a straight trajectory can be qualitatively understood as follows. When the shape becomes asymmetric, the net force shifts further away from the centre-of-mass, leading to a stronger torque. Beyond a certain threshold value of 2b, this torque becomes very large, and the particle exhibits a nearly pivoted rotational motion. We observe that the direction of rotation of the elliptical Janus particle is random, as shown in Fig.\ref{fig:1} (c), (d). However, once the particle begins to rotate, it maintains its rotational direction until it collides with a boundary.

\section{Numerical model}
\subsection{\textbf{Model Description}}
To study the observed rotational dynamics of the elliptical Janus particle on the surface of water due to Marangoni forces, we follow the two-dimensional framework proposed by Kitahata et al.~\cite{kitahata}. We assume that the  concentration field \( c(\vb{r}, t) \) of the camphor on the surface evolves according to the equation,

\begin{equation}
\label{eq:react_diff}
\frac{\partial c(\vb{r}, t)}{\partial t} = D \nabla^2 c(\vb{r}, t) - \alpha c(\vb{r}, t) + f(\vb{r_c}(t),\chi(t)),
\end{equation}
 \( D \) is the diffusion coefficient of camphor on the surface, \( \alpha \) is the degradation rate of camphor molecules, and $f(\vb{r_c}, \chi)$ denotes the source term. In this context, the source term indicates the camphor-coated half of the Janus particle, where
 , \( \vb{r_c}(t) \) and \( \chi(t) \) denote the centre-of-mass position and orientation of the particle.

The surface tension \( \gamma \) is assumed to decrease linearly with the local camphor concentration $c$,
\begin{equation}
\gamma(c, \vb r, t) = \gamma_0 - \kappa c(\vb r, t),
\label{surface_tension}
\end{equation}
where \( \gamma_0 \) is the surface tension of pure water, and \( \kappa \) is a positive constant. The particle experiences a net Marangoni force \( \vb{F} \)  arising from surface tension gradients along its boundary \( \Omega_{\vb r_c} \) is given by the relation,

\begin{align}\label{force_equation}
\vb{F} &= \oint_{\Omega_{\vb r_c} } \gamma(c) \vb{n} \, ds.
\end{align}
Since we are studying the dynamics of Janus particles, the torque \( \bm{\tau} \)  about ${\bf r}_c$ is separately calculated,
\begin{align}
\bm{\tau} &= \oint_{\Omega_{\vb r_c} } \left(\vb{r} - \vb{r_c}(t)\right) \times \gamma(c) \vb{n} \, ds,
\label{torque_equation}
\end{align}
where \( \vb{n} \) is the outward normal on the article boundary. These relations qualitatively capture force and torque due to the Marangoni effect. The net force and torque calculated for the particle are incorporated in the dynamical equations of particle position and alignment,
\begin{align}
m \frac{d^2 \vb r_c}{dt^2} &= -\eta_{\parallel} \vb v_{\parallel} - \eta_{\perp} \vb v_{\perp}+\vb{F},
\label{translation}\\
I \frac{d^2\chi}{dt^2} &= -\eta_r \frac{d\vb{\chi}}{dt} +  \bm{\tau}
\label{orientation}
\end{align}
Due to the aniostropic shape of the particle, it is necessary to consider the parallel and perpendicular components of the velocity, $\vb v_{\parallel}$ and $\vb v_{\perp}$, such that, the net velocity of the particle $\frac{d \vb r_c}{dt} = \vb v = \vb v_{\parallel} + \vb v_{\perp}$. Also the angular velocity $\bm{\omega} = \frac{d\chi}{dt}$. Since we study the dynamics of a single particle, long-range, non-local hydrodynamic interactions and the evolution of the surrounding flow field are not included in the model. The effects of local hydrodynamics are captured by the local drag coefficients, $\eta_{\parallel}$, $\eta_{\perp}$ \, and $\eta_r$, which are the parallel, perpendicular, and rotational drag coefficients, respectively.

As observed in the experiments, the type of trajectories displayed by the particle depends on the particle-shape. One of the reasons for the dependence is that the distribution of forces and torques due to the Marangoni effect depends on the shape of particle contour, as evident in the equations~\ref{force_equation}-~\ref{torque_equation}.    
Another important aspect that determines the trajectory is the anisotropic drag experienced by the particle on the water surface due to its shape. However, an exact estimation of these drag coefficients for partially submerged particles is challenging~\cite{sur2021effect, ooi2016measuring, wilt2024activecheerios,kang2021marangoni, Boniello2015, Boniello2016}. On the other hand, neglecting the anisotropy in drag will lead to an incorrect estimation of particle trajectories~\cite{Kitahata2}. Various combinations of translational and rotational drag give rise to different trajectories, like stationary states, rotational modes, and motion along the short axis. So, a proper relationship between these drag coefficients with the particle's shape and size needs to be established to capture the experimental observations correctly.

In this study, we adopt a simplified approach by estimating the drag coefficients for a spheroid immersed in a bulk fluid, characterized by its major and minor axes, $a$ and $b$, respectively. This approach follows previous theoretical~\cite{perin} and experimental~\cite{MUKHIJA200798} studies on circular and rod-like particles. Although the long-ranged hydrodynamic behaviour is qualitatively different for 2D and 3D systems, in `dry' active matter systems such bulk drag coefficients are employed in 2D simulations that only consider local viscous drag. Moreover, the drag coefficient for spheroids captures the orientation dependance of local translational and rotational drag and qualitatively reproduce the experimental observation. 
The drag coefficient $\eta$ for a spheroid is split into parallel $\eta_{\parallel}$ ( parallel to the major axis), perpendicular $\eta_{\perp}$ (perpendicular to the major axis), and rotational $\eta_r$ (rotation of the body axis about the centre-of-mass) such that,
\begin{align}
\label{t_drag}
    \eta_{\parallel or \perp} &= 6 \pi \zeta b G_{\parallel or \perp} \\
    \eta_{r} &= \zeta V_0 G_r
    \label{rotation}
\end{align}
Here, $a$ and $b$ are the semi-major and semi-minor axes, respectively, and $V_0$ is the volume of the particle, which for a spheroid is given by $V_0 = \frac{4}{3} \pi ab^2$ and $\zeta$ denote the effective viscosity of the medium For the elliptical particle, we would stick to the prolate spheroid for which the Perin friction factors are given by,
\begin{align*}
G_{T}^{\parallel} &= \frac{4}{3}\left[\frac{p}{(1-p^{2})} + \frac{2p^{2}-1}{(p^{2}-1)^{3/2}}\ln(p + \sqrt{p^{2}-1})\right]^{-1} \\
G_{T}^{\perp} &= \frac{8}{3}\left[\frac{p}{(p^{2}-1)} + \frac{2p^{2}-3}{(p^{2}-1)^{3/2}}\ln(p + \sqrt{p^{2}-1})\right]^{-1} \\
G_{\theta} &= \frac{2}{3}\frac{(p^{4}-1)}{p}\left[\frac{(2p^{2}-1)}{\sqrt{p^{2}-1}}\ln(p + \sqrt{p^{2}-1}) - p\right]^{-1}
\end{align*}

Putting this expression for translational and rotational drag coefficients in equations~\ref{translation}-~\ref{orientation} completes the description of the numerical model used. 

To simulate the system, the particle is initially kept at the centre of a square box with periodic boundary conditions, with grid spacing $dx,dy = 0.05a$. The particle releases camphor from its body, which has a degradation rate $\alpha$. In simulations to specifically study the effect of particle shape, we keep $\alpha = \alpha_0$, which is kept as the unit of inverse time. The camphor diffusion coefficient $D = 25 a \alpha_0^2$, and the system evolves with a $\Delta t = 2 \times 10^{-4} {\alpha_0}^{-1}$. For the current analysis of varying particle shape, $\kappa$ is set to $0.2 \frac{\gamma_0 f_0}{\alpha_0}$. The instantaneous camphor concentration $c$ is calculated numerically by integrating the discretised version of the equation~\ref {eq:react_diff} using the Euler's method. Subsequently, the reduction in surface tension $\gamma$ and the instantaneous force and torqe on the particle is calculated. Finally, the particle position and orientation are updated by numerically solving equations~\ref{translation}
and ~\ref{orientation} using Euler's method. Simulations for Janus particles were conducted by systematically varying the particle shape, degradation rate $\alpha$ and the effective Marangoni strength, $\kappa$. Each set of parameter values was simulated up to seven different initial realisations. 

\begin{figure}
    \centering
    \includegraphics[width=1\linewidth]{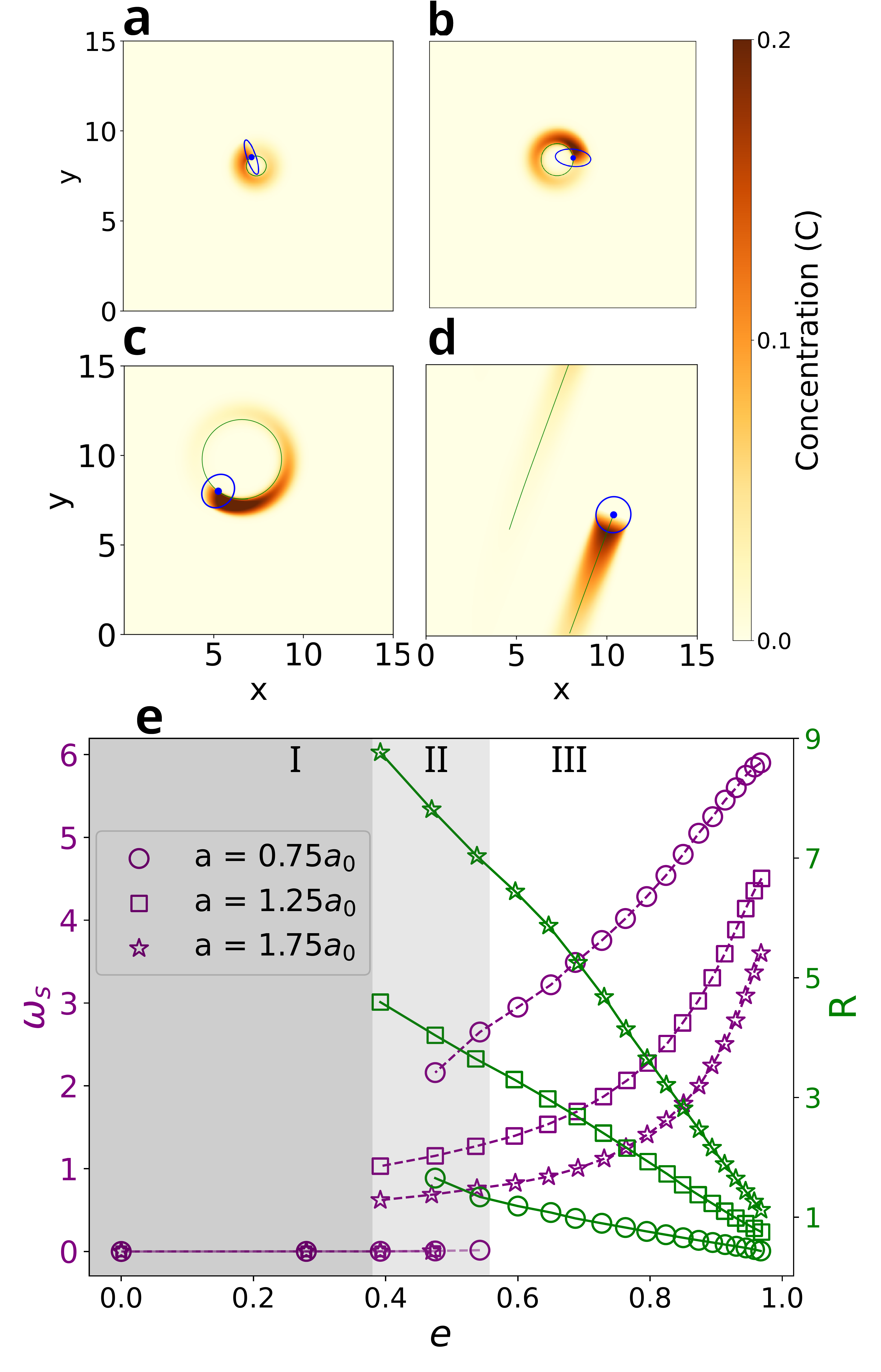}
    \caption{(a-d) shows 4 different shapes from circle to elongated ellipse, depicting distinct trajectories from straight-line to circle (also see Movie2).Fig. (e) shows how the angular velocity $\omega$ and the radius $R$ change upon changing the shape of the particle. For this case, $\alpha = \alpha_0,~$ and $\kappa = 0.2 \frac{\gamma_0 f_0}{\alpha_0}$}
    \label{fig:sim1}
\end{figure}
\subsection{\textbf{Numerical results}}
\begin{figure*}
    \includegraphics[width=1\linewidth]{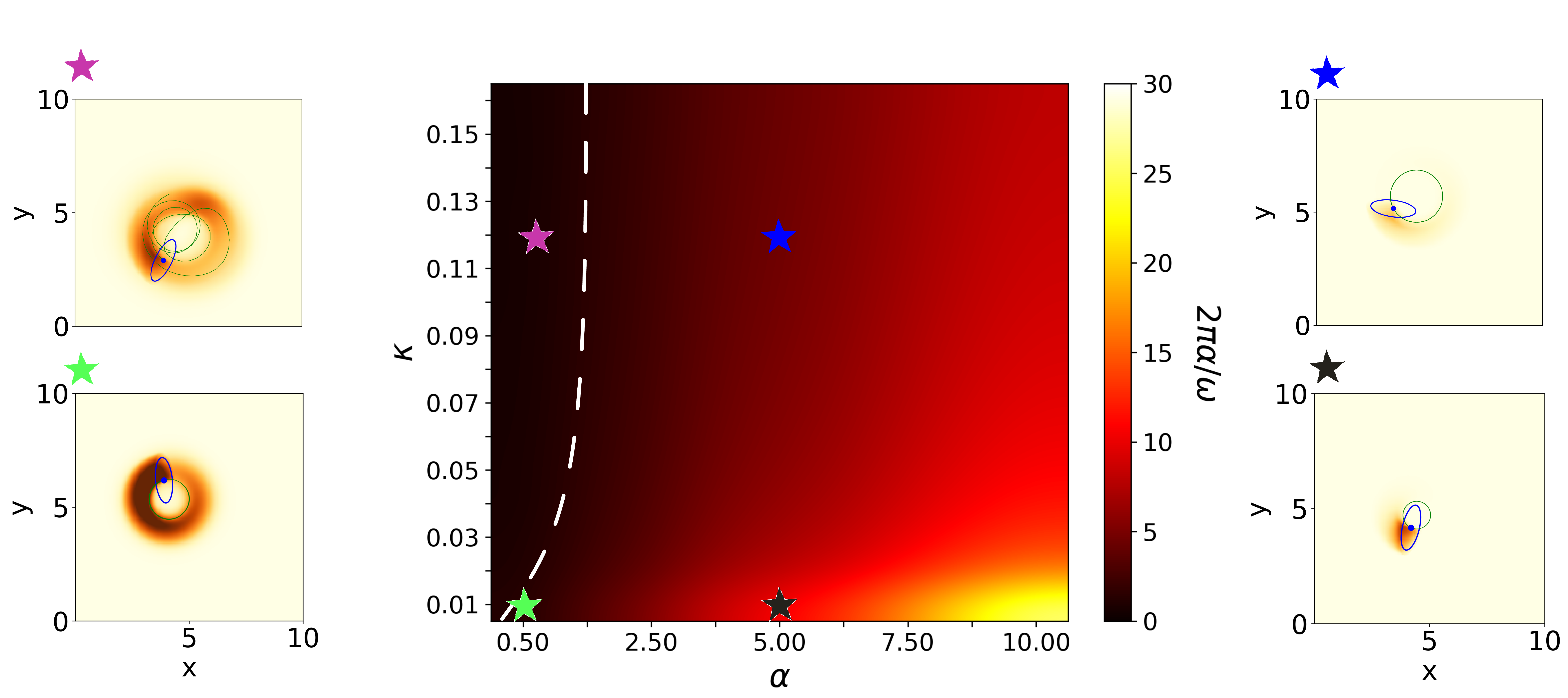}
    \caption{The figure shows that the $\alpha, \kappa$ determine the trajectory of the particle of eccentricity $e=0.93 ~~(a = 1.25~a_0 ,~b=0.46~a_0)$ in the steady state (in units of $\alpha_0$ and $\kappa_0 = \gamma_0 f_0 / \alpha_0$). The white-dotted curve gives a rough boundary for the ratio $2\pi\alpha/\omega$ near 1. When this happens, the circular trajectory is disturbed, and the particle deviates to a complex trajectory.}
    \label{fig:sim2}
\end{figure*}
\noindent Initially, the coupled evolution of the camphor concentration field and the particle dynamics was simulated for an elliptical particle which is completely coated with camphor. The simulation has successfully reproduced the previously established result \cite{kitahata} and verified that when the particle is fully camphor-coated, any initial torque generated by asymmetries in the camphor distribution decays over time. As a result, the system invariably relaxes to a state of persistent translational motion along the short axis, which represents the stable steady state for such particles.

\subsubsection{Effect of particle shape}
 Having established this baseline, we next turned to the case of Janus particles, in which only one-half of the particle is coated with camphor. First, we systematically investigated how the particle's trajectory depends on its shape. To probe the effect of geometry, we varied the eccentricity $e$ of the elliptical particle by fixing the semi-major axis $a$ and gradually decreasing the semi-minor axis $b$ from $a$ down to $0$. The limit $a=b$ corresponds to a circular particle ($e=0$). In this isotropic case, our simulations show that the only stable trajectory is a straight line, characterized by a vanishing angular velocity $\omega = 0$ [Fig.~\ref{fig:sim1}(e), region I, and Fig.~\ref{fig:sim1}(d)]. This result is consistent with the experimental observation.

\noindent As the eccentricity $e$ is increased, however, a qualitatively different behavior emerges. Beyond a certain critical value of $e$—which itself depends on the absolute size of the particle $a$—a new stable solution arises. In this case, the particle executes a circular trajectory with a finite angular velocity $\omega = \omega_{s}$ (Fig.~\ref{fig:sim1}(a)-(c) and region III in Fig~\ref{fig:sim1}(e)). Close to this transition point from a straight to a circular trajectory, the system passes through an intermediate bistable regime (region II in Fig~\ref{fig:sim1}(e)), where both the straight-line solution ($\omega=0$) and the circular solution ($\omega=\omega_{s}$) are stable for different runs. In this bistable window, the outcome 
depends sensitively on initial conditions and initial perturbations. The system may settle into either persistent translation or sustained rotation for the same parameter values. In Fig.~\ref{fig:sim1}(e), we compute the steady-state angular velocity $\omega_{s}$ and the radius of the circular trajectory, $R$, as a function of $e$ and for three different values of $a$. As shown in Fig.~\ref{fig:sim1}(e),  within region III, the angular velocity $\omega$ increases monotonically with eccentricity $e$, reflecting the fact that the torque experienced by the particle grows with the degree of asymmetry in the underlying camphor 
distribution. In addition to $\omega$, the radius of the trajectory $R$ also reaches a steady state. Within the bistable region (region-II in 
Fig~\ref{fig:sim1}(e)), we observe that for small $e$, the circular trajectories are comparatively large in radius (Fig.~\ref{fig:sim1}(c)), whereas increasing $e$ systematically reduces the radius, as illustrated in Figs.~\ref{fig:sim1}(a)--(c). This trend is followed for all values of $a$ (Fig~\ref{fig:sim1}(e)) and they are consistent with the trend observed in the experiments (Fig~\ref{fig:2}(f)). It should also be noted that larger sized particles (with larger $a$) show a smaller $\omega_c$ and higher $R$ for the same value of $e$, as evident in Fig~\ref{fig:sim1}(e). 

Both the experiments and the simulations show that an increase in particle eccentricity $e$ leads to an increase in $\omega$ and a decrease in $R$. This observation can be rationalized by considering the particle geometry. As the eccentricity increases (achieved here by reducing the semi-minor axis $b$ while keeping the semi-major axis $a$ fixed), the perimeter of the camphor-coated region shrinks. This reduction 
lowers the net driving force $F_{\text{s}}$, according to Eq.~\ref{force_equation}. At the same time, the enhanced asymmetry in camphor release associated with larger $e$ sustains a comparatively stronger torque $\tau_{\text{s}}$, since the decay of torque with eccentricity is slower than that of the force. Since $F_S$ determines the steady linear velocity $v_s$ and $\tau_s$ determines the angular velocity $\omega_s$, this effect is reflected as an increase in $\omega_s$ with $e$. Further, the steady-state trajectory radius is determined by the ratio $R = \frac{v_{\text{s}}}{\omega_{\text{s}}}$, also increases as a result. In the asymptotic limit of highly elongated particles, the radius approaches a finite lower bound of $R \approx a/2$.

\begin{figure}
    \centering
    \includegraphics[width=1\linewidth]{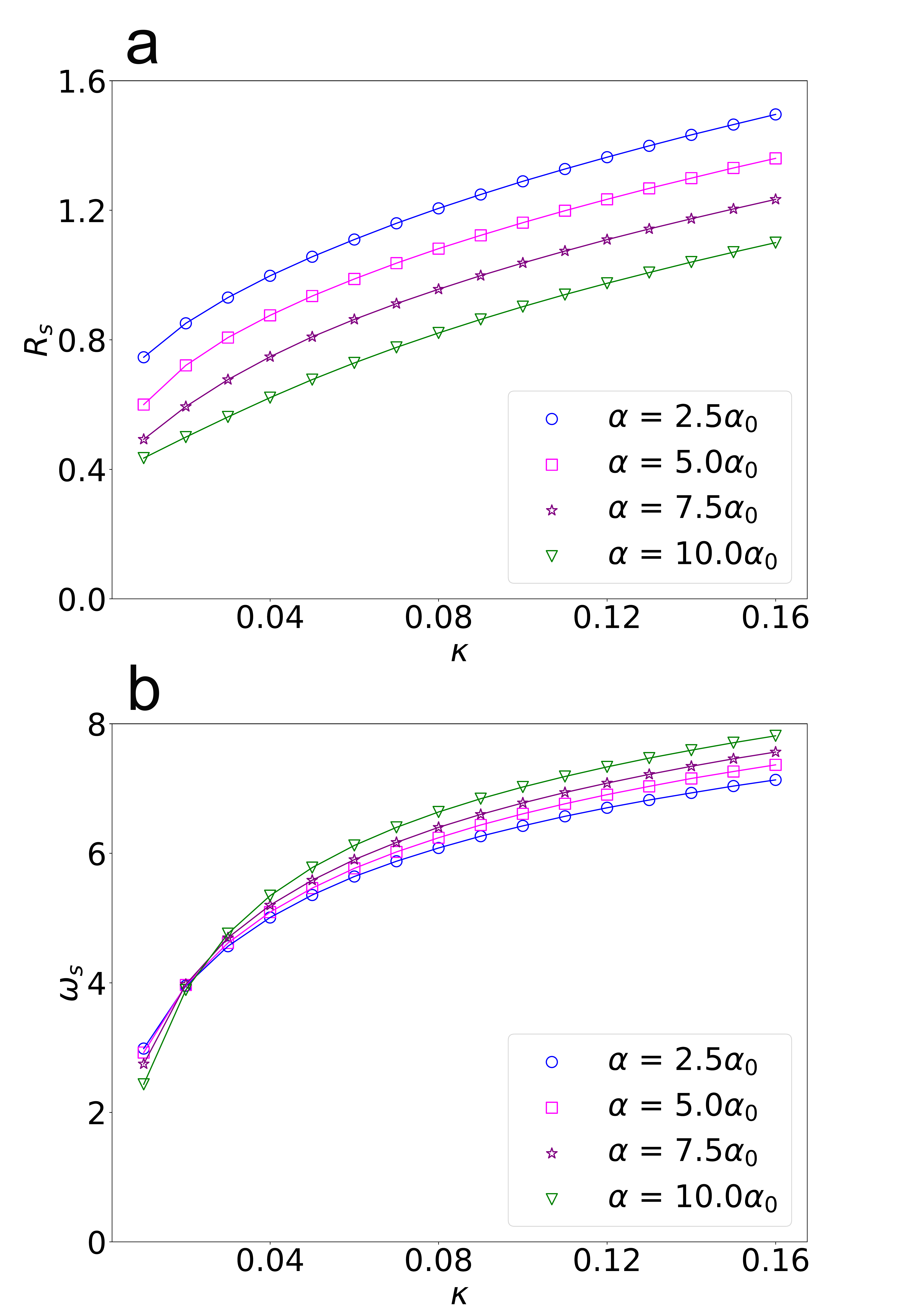}
    \caption{The radius $(R)$ and $\omega$ of the particle is plotted by changing $\alpha$ and $\kappa$. Since the surface tension dip due to the surfactant depends on $\kappa$, increasing it increases both the radius $R$ and $\omega$.}
    \label{fig:sim3}
\end{figure}
\subsubsection{Effect of chemical properties}
Regarding the surfactant field, $\alpha$ represents the degradation (sublimation) rate of the surfactant, while $\kappa$ quantifies the reduction in water surface tension caused by the surfactant. Both parameters play a crucial role in determining the type of dynamics exhibited by the particle. At this stage, it is useful to introduce certain characteristic length scales of the system. For a stationary particle, the spread of camphor governed by the reaction–diffusion equation~\eqref{eq:react_diff} is given by $l_1 = \sqrt{\tfrac{D}{\alpha}}$. However, due to initial perturbations, the particle begins to move with a speed $v$, which eventually reaches a steady constant value.
As discussed earlier, the steady-state speed $v_{s}$ depends on the particle size, geometry, and the parameters $\alpha$ and $\kappa$. While moving at speed $v_{s}$, the particle leaves behind a surfactant trail of characteristic length $l_2 = \tfrac{v_{s}}{\alpha}$ [see Fig.~\ref{fig:sim1}(d)]. Owing to its asymmetry, the particle follows a curved trajectory, whose radius $R$ is determined by the ratio of $F_s$ to $\tau_s$. Tracing this curved path, the particle eventually returns to its initial position, thereby forming a circular trail. If the trail length is shorter than the circumference of the circle $2 \pi R$, the particle encounters no camphor trace at the starting point and thus continues along the same circular path. In contrast, if the trail length is comparable to or larger than the circumference, the particle experiences a modified surfactant field after one revolution, causing it to deviate from the circular trajectory. So we get a condition for an unperturbed 
    $2 \pi R \gg l_2 $ or $ \frac{2 \pi \alpha }{\omega} \gg 1 \nonumber$. 
In Fig.~\ref{fig:sim2} we plot a phase diagram for a fixed $e$ in the $\kappa -\alpha$ space. The color code indicates the dimensionless parameter $2\pi\alpha/\omega$. The dotted white curve indicates $2\pi\alpha/\omega =1$ and it roughly separates the darker region in the phase diagram where the two length scales ($2\pi R$ and $\l_2$) become comparable and the particle deviates from the circular trajectory.

In addition to the change in radius of the circular trajectory, both $\alpha$ and $\kappa$ change the surface tension and thereby the Marangoni force and torque acting on the particle. For a given value of $e$, $\omega_s$ and $R_s$ are determined by $\alpha$ and $\kappa$ via Eq~\ref{surface_tension}-~\ref{orientation}. In Fig~\ref{fig:sim3}, we show how $R_s$ and $\omega_s$ changes as a function of $\kappa$ for different $\alpha$. Our analysis shows that the radius of the trajectory monotonously increases with $\kappa$  and an increase in $\alpha$ causes a decrease in $R_s$ for all values of $\kappa$ (Fig~\ref{fig:sim3}(a)). The angular velocity $\omega_c$ also increases monotonously with $\kappa$. However, the dependence on $\alpha$ is more prominent for higher $\kappa$, as shown in Fig.~\ref{fig:sim3}(b).


\section{Conclusions}
In this work, we have experimentally studied the self-propulsion of an elliptical Janus camphor particle at the air–water interface and examined how its shape controls its motion. Experiments show a straight trajectory for circular Janus particles, and a circular trajectory for elliptical Janus particles. Experiments also reveal that as the particle become more elongated, the radius of the circular trajectory decreases and its angular velocity increases.

The experimental findings are rationalized using simulations that reveal distinct dynamical regimes that emerge as both particle geometry and surfactant parameters are varied and it qualitatively reproduces the experimental observations.
 As the particle becomes more elongated, a bistable regime appears: depending on initial conditions, the particle either moves straight or follows a circular path with a well-defined angular velocity. At larger eccentricities, the bistability disappears, and the particle reliably moves in circles, with angular velocity increasing as eccentricity increases. The radius of this circular motion decreases with eccentricity because the camphor-coated perimeter becomes smaller, reducing the net propulsive force faster than the torque. This imbalance—weakening force but comparatively persistent torque—produces tighter circular paths for more elongated particles. Finally, comparing the trail length with the curvature radius clarifies when stable circular trajectories occur. This leads to a simple division of the $(\alpha,\kappa)$ parameter space into regions that support clean circular motion and those where the particle deviates from it.

We note that the numerical model does not explicitly consider the fluid velocity or flow-induced mixing of camphor. Despite these simplifications, 
our simulations qualitatively reproduce the experimentally observed trajectories, demonstrating that the interplay between the surfactant field and particle geometry is sufficient to explain the behaviors. The hydrodynamic effects due to the wall are not important as long as the particle is far from the wall of the petri dish. These results highlight how geometric asymmetry amplifies or suppresses chemo-mechanical feedback, thereby enabling or destabilizing specific modes of motion. By systematically varying geometric parameters such as eccentricity and aspect ratio, we show that particle shape directly modifies the balance of Marangoni forces and torques, thereby selecting or switching between distinct modes of motion. As a next step, one can study the collective dynamics of multiple Janus particles to understand the type of interactions and their effect on collective motion. The interparticle interactions in such systems are expected to be  more complex and challenging to model than those in typical dry active matter, as they can be influenced by both fluid flows and capillary effects. In addition, connecting such particles into chains can also reveal novel dynamical states~\cite{sadhu2025active, ishikawa2022pairing, ishikawa2025simple}.

The study opens pathways to control Marangoni-driven particles solely by adjusting their properties, such as shape, size, and surfactant nature and concentration. Similar attempts are made with particles by tuning other environmental properties such as airflow, evaporation gradients, confinement, and surface-tension landscapes to guide trajectories of these systems and achieve effective external control. By tuning these environmental fields, one can guide trajectories, switch dynamical modes \cite{ionic}, or even trap particles within localized regions \cite{capture2023Tiwari}.
 
Beyond offering insight into fundamental active matter physics, our findings may inform the design of future self-propelled micromachines, environmental sensors, and soft-robotic elements that operate under autonomous, fuel-driven, and symmetry-broken conditions.
\begin{appendix}
    \section{Supplemantary movies}
    \begin{enumerate}
        \item {\bf Movie1}: The effect of the shape of the particle -- comparison between experiment and theory.
        \item {\bf Movie2}: Simulation movie showing the effect of $\alpha$ and $\kappa$ for a fixed eccentricity $e=0.93$.
    \end{enumerate}
\end{appendix}
\bibliography{reference}


\end{document}